\documentclass{article}
\usepackage{ijcai23}

\usepackage{float}
\usepackage{times}
\usepackage{wrapfig}
\usepackage{soul}
\usepackage{url}
\usepackage[hidelinks]{hyperref}
\usepackage[utf8]{inputenc}
\usepackage[small]{caption}
\usepackage{graphicx}
\usepackage{amsmath}
\usepackage{amsthm}
\usepackage{booktabs}
\usepackage{multirow}
\usepackage{algorithm}
\usepackage{algorithmic}
\usepackage[switch]{lineno}
\usepackage{array}
\usepackage{caption}            
\usepackage{subcaption}         
\usepackage{amsfonts}           
\usepackage{duckuments}         
\usepackage{tabularx}
\usepackage{paralist}
\usepackage{threeparttable}

\title{Link Prediction for Social Networks using Representation Learning and Heuristic-based Features}

\author{
    Samarth Khanna$^{*}$
    \and
    Sree Bhattacharyya$^{*}$
    \and 
    Sudipto Ghosh
    \and
    Kushagra Agarwal
    \And
    Asit Kumar Das
    \\
    \affiliations
    Indian Institute of Engineering Science and Technology, Shibpur
    \\
}

\begin{document}

\maketitle

\def\thefootnote{*}\footnotetext{Equally contributing authors.}
\begin{abstract}
    The exponential growth in scale and relevance of social networks enable them to provide expansive insights. Predicting missing links in social networks efficiently can help in various modern-day business applications ranging from generating recommendations to influence analysis. Several categories of solutions exist for the same. Here, we explore various feature extraction techniques to generate representations of nodes and edges in a social network that allow us to predict missing links. We compare the results of using ten feature extraction techniques categorized across Structural embeddings, Neighborhood-based embeddings, Graph Neural Networks, and Graph Heuristics, followed by modeling with ensemble classifiers and custom Neural Networks. Further, we propose combining heuristic-based features and learned representations that demonstrate improved performance for the link prediction task on social network datasets. Using this method to generate accurate recommendations for many applications is a matter of further study that appears very promising. The code for all the experiments has been made public$^1$\def\thefootnote{1}\footnotetext{\href{https://github.com/SamarthKhanna/social-network-link-prediction}{Source Code}}.
\end{abstract}

\begin{table*}
	\centering
	\begin{tabular}{p{0.05\textwidth}p{0.07\textwidth}p{0.07\textwidth}p{0.07\textwidth}p{0.07\textwidth}p{0.07\textwidth}p{0.07\textwidth}p{0.07\textwidth}p{0.07\textwidth}p{0.07\textwidth}}
	    \toprule
	   \textbf{Metric} & \textbf{soc-Epinions} & \textbf{GitHub} & \textbf{Twitch-DE} & \textbf{Twitch-EN} & \textbf{Twitch ES} & \textbf{Twitch-FR} & \textbf{Twitch-PT} & \textbf{Twitch-RU}  & \textbf{wiki-Vote}\\
	   \midrule
	   \textbf{Nodes} & 75,879 & 37,700 & 9,498 & 7,126 & 4,648 & 6,549 & 1,912 & 4,385 & 7,115 \\
	   \textbf{Edges} & 508,837 & 289,003 & 153,138 & 35,324 & 59,382 & 112,666 & 31,299 & 37,304 & 103,689 \\
	   \bottomrule
	\end{tabular}
	\caption{Dataset Statistics: A count of the Nodes and Edges present in each network}
        \label{Table 1}
\end{table*}

\section{Introduction}

With the tremendous boost in the popularity and use of social media in recent times, it is imperative to devise efficient methods to mine information from social networks. Link prediction (specifically on social networks) \cite{liben2007link} is one such area. In a social network, nodes represent the users or entities interacting in a public forum, and the edges denote the interactions between nodes. The task of link prediction is to predict whether a relationship (or an edge) \textit{can} exist between two nodes in a given network. This can be approached in several ways considering several types of networks. Most existing methods depend on the network structure or on attributes of nodes and edges in the network. One of the primary tasks in predicting links is generating features to represent the network's nodes and edges. These features can then be used to model the relationships between entities in a social network using classical machine learning or neural network-based classifiers. Several methods exist for engineering such features which are analyzed in this work. 

The first broad category explored is based on learning graph representations \cite{replearning1} \cite{replearning2} for representing nodes in social networks. Essentially it means that nodes in a graph are projected as low-dimensional vectors in a latent space while preserving the graph properties. 
Under this, the methods studied include graph structural embeddings \cite{graphwave}, neighborhood-based embeddings \cite{neigh-emb}, and Graph Neural Networks \cite{gnnreview}. 
The second category studied comprises a relatively simpler method that utilizes the global topological structure of the graph. These include degree-related features, similarity-based features, ranking features like PageRank \cite{pagerank}, graph-based features like the Adamic-Adar index \cite{adamicAdar}, and preferential attachment features. 

Our primary experiment aims to understand how different methods, under the above two broad paradigms, perform in comparison with each other, on a popular data set \cite{soc-epinions} to begin with. Following the extraction of features, the link prediction problem on social networks has been formulated as that of binary classification. This means, given the features of an unseen edge, the task is to predict whether it should belong in the network or not. Ensemble learning techniques such as XGBoost \cite{xgboost}, RandomForest \cite{randomforest}, and LightGBM \cite{lightgbm}, as well as custom Neural Networks are employed for this purpose. Although, while creating training and test sets, the data splits generated are kept balanced in terms of the presence of positive and negative edges, the F1 score is selected as the final metric since it eliminates any possibility of misleading conclusions in case there is a bias towards one class.

Based on the results arising from the previous step, the better-performing algorithms are tested on multiple data sets with subtle differences in their characteristics. Once a uniform performance trend is detected in subsequent findings, further experiments are carried out to determine whether a combination of the best-performing features would lead to an improvement in results. In particular, an attempt is made to combine heuristic-based features with the representation learning methods to determine whether it boosts link prediction performance. The outcomes of this experiment are promising. Various configurations of custom Deep Neural Networks have also been introduced which yield better performance than the ensemble learning techniques used. 
The main contributions of this work entail exploring various combinations of interpretable heuristic-based features and learnable black-box features. Secondly, custom Neural Network architectures for yielding improvements in performance over tree-based ensemble models are also introduced. The results obtained encourage deeper research into how deterministic and learned features complement each other. Ways in which they can be interweaved to produce results that neither paradigm of learning can individually achieve are a matter of further exploration.

\section{Data Used}

\label{D}
A total of four datasets from the Stanford Network Analysis Project (SNAP) \cite{leskovec2014snap} are considered, which contain various kinds of relationships between users of online platforms. The datasets used are:
\begin{itemize}
\addtolength\itemsep{-0.17cm}
    \item \textbf{Epinions Social Network} \cite{soc-epinions} (directed): This is a trust-based, medium-sized social network, used by several other related works for a variety of tasks \cite{epinions-use1}\cite{epinions-use2}\cite{epinions-use3}. An edge in the Epinions network represents a relationship in which a user chooses to trust another based on the latter's published reviews on the website.
    \item \textbf{GitHub Social Network} \cite{musae} (undirected): This network provides information on "follow" relationships between users having a minimum of 10 starred repositories. 
    \item \textbf{Wikipedia Vote Network} \cite{wiki-vote1} \cite{wiki-vote2} (directed): This is a network that contains data about voting pertaining to Request for Adminship (RfA) on Wikipedia.
    \item \textbf{Twitch Social Networks} \cite{musae} (undirected): These represent friendships between gamers who stream in a certain language, with there being a separate sub-network for each such language.
\end{itemize}
None of these graphs are weighted or have any edge features. While the extent of generalizability of the results across networks with similar properties of trust, reliability, and influence has not been explored in this work beyond the scope of the selected data sets, extrapolating the observations to and gaining insights from data such as scientist collaboration graphs and YouTube subscriber networks can be studied further. Table \ref{Table 1} contains the count of nodes and edges for each graph used.

\section{Methodology}
\begin{figure*}
    \centering 
    \includegraphics[scale=0.45]{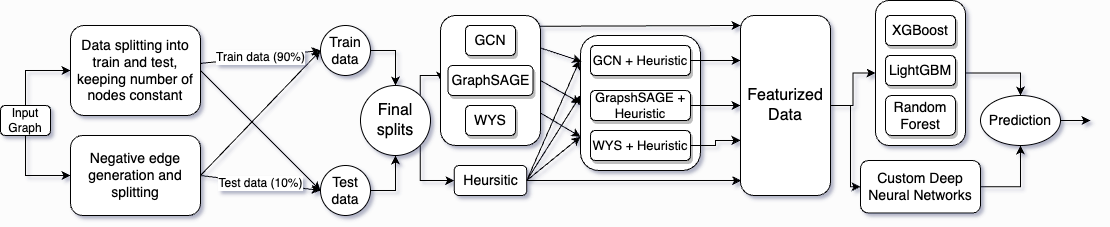}
    \caption{Visualization of the workflow for adopted in our methodology for each data set. }
    \label{fig:1}
\end{figure*}

The entire pipeline of the methodology used is described in this section. Figure \ref{fig:1} provides an overview of the workflow.

\subsection{Data Preparation}
The link prediction problem is approached as a binary classification task. The edges already present in the graph belong to the 'positive' class and the rest belong to the 'negative' one.    

Splitting graph datasets into training and testing data is often a tricky task. A popular strategy employed is removing the edges to be used for testing from the training data only if the number of nodes and the number of connected components remained unchanged in the training graph by doing so. This ensures that all test edges are relationships between previously seen nodes, and no information is lost by breaking an existing connected component. However, this method proves to be very time-consuming as it involves checking for reduction in the number of connected components, every time an edge is removed. The other, much simpler alternative is to split the edges randomly into training and testing data. Although that consumes significantly less time, it leads to chances that a node absent in the training set might appear in the test set. Some of the feature generation methods used, however, have a hard requirement that all nodes in the test graph also be present in the training graph. Hence, a middle ground is adopted and the splits are made ensuring just that the number of nodes in the training graph does not reduce. This substantially speeds up the process, while also avoiding the problem of unseen nodes. 

For the negative edges, splits are simply created that mimic the proportions of the positive edge data. As the number of potential negative edges is of the order of $n^2$ (\(n =\) number of nodes), a subset containing randomly selected edges is created. A filter for candidate negative edges is that the shortest path between the two involved nodes in the original graph is at least 3. The number of such edges chosen is roughly the same as the number of positive edges, in order to get a balanced data set.

The final splits of data used (for each data set) are as follows:
\begin{itemize}
\addtolength\itemsep{-0.17cm}
    \item Training data: 90\% of positive edges + 90\% of negative edges 
    \item Testing data: 10\% of positive edges + 10\% of negative edges 
\end{itemize}
Before using this data for training and testing respectively for link prediction, the representative features for each data point (edge) are generated using the techniques discussed below. Wherever training of the representation-generating algorithms is involved, the algorithm is fit only on the training graph. 

\subsection{Feature Generation}

The primary focus of our study is presented in this section, detailing the different methods used for generating node and edge representations. The usage of these particular algorithms for obtaining node embeddings was suitable as it provided a diverse base of paradigms to compare against. 

\subsubsection{Structural Node Embeddings}
The algorithm used that falls under this category is GraphWave \cite{graphwave}. It is based on unsupervised training and does not depend on any node features or node labels. It finds structural representations of nodes based on structural similarities in the network topology. It represents each node as a low-dimensional embedding by considering the diffusion pattern of a spectral graph wavelet centered at each node. Two nodes in disconnected parts of a graph can have a very similar structural node embedding if their local neighborhoods have similar topology. To implement this, the StellarGraph framework was used \cite{StellarGraph}.

\subsubsection{Neighborhood-based Node Embeddings}
Under this category, algorithms focus on relationships between nodes, as against local graph structure. Two nodes that are related or close to each other in a network will be assigned similar representations in the low-dimensional embedding space by this category of methods. The algorithms used that fall under this broad description are Node2Vec \cite{node2vec}, DeepWalk \cite{Deepwalk}, Walklets \cite{walklets}, NetMF \cite{netmf}, and NodeSketch \cite{Nodesketch}. Further, the algorithms can be subdivided into those employing Matrix Factorization and those employing Random Walks \cite{neigh-emb}. The definition of whether two nodes are "close" to each other in a network varies in methods using Matrix Factorization and Random Walks. Algorithms based on Matrix Factorization perform low-rank approximation on the adjacency matrix of the network that captures its global topology. NetMF \cite{netmf} falls under this subdivision. On the other hand, Random Walk-based methods, inspired by Word2Vec \cite{node2vec} for word representations, sample neighbors of a node using random walks. This creates the context of a node.
Then the random walk sequences are fed into a SkipGram encoder \cite{skipgram} that learns representations for the nodes based on the similar nodes sampled. DeepWalk \cite{Deepwalk} and Node2Vec \cite{node2vec} use Random Walks. DeepWalk, however, uses only local structural information to determine which nodes are similar, through truncated random walks. Node2Vec improves this by using biased random walks to incorporate global topological information of the network. Walklets improve on DeepWalk by using random walks that can skip or hop multiple neighbors at a time \cite{walklets-review}. Finally, NodeSketch works using recursive sketching and also preserves high-order node proximity. All the algorithms under this category are implemented using the Karateclub library.

\subsubsection{Graph Neural Network-based Node Embeddings}
This class of methods aims to extract higher levels of features from graphs in an automated paradigm through weight optimization. Graph Convolutional Networks (GCNs) \cite{gcn} are primarily a generalization of Convolutional Neural Networks (CNNs) \cite{CNN} in the graph domain. It introduces parameter sharing across graph nodes. Every layer of the network learns latent representations to encode graph nodes, by aggregating information about the graph structure successively \cite{gnnreview}. The encoding thus obtained from the last layer can be used as node embeddings or features for any predictive task. The three algorithms we use under this category are GCN, GraphSAGE \cite{graphsage}, and Watch Your Step.

GCN is one of the most popular variants of a Graph Neural Network (GNN) \cite{gnnreview}. The key difference between GCN and CNN, which it is inspired by, is that in the case of graphs the size of the receptive fields and filters cannot be as easily determined. GCNs deal with this by sharing weights for all n-hop neighbors. Through the simple convolution operation, GCNs aggregate information through its layers, where each layer considers progressively higher orders of neighbors of a node. In our study, GCN trains using the Deep Graph Infomax approach \cite{GraphInfomax}. 

GraphSAGE \cite{graphsage}, a deep inductive network, encodes a node's attribute level information to obtain its embedding. Being an inductive model, once trained, it can generate node embeddings for a previously unseen network and thus forms an ideal use case for resource-poor scenarios. It also improves upon the GCN by adding more aggregators like Pooling and LSTM. \\
The Watch Your Step \cite{wys} algorithm, on the other hand, is based on Graph Attention \cite{gatsurvey}. It makes hyperparameters of the deep learning-based models (such as the length of the random walk, and the length of node2vec) learnable. It works by enforcing the model to pay attention to certain parameters to optimize an upstream objective.

Out of the three algorithms studied, GraphSAGE and GCN also additionally use node-level attribute information. In both cases, the 3 node features passed as attributes are Katz Centrality \cite{katz}, PageRank \cite{pagerank}, and node degree. All three algorithms are implemented using the StellarGraph framework \cite{StellarGraph}.

\subsubsection{Heuristic-based Methods}
\label{HBM}
The final paradigm of feature extraction leverages deterministic properties of nodes and edges that can be inferred from the graph. 
Given below are the broad categories of features we have used, the number of features in each category, and the exact features used. PageRank, Katz Centrality, and HITS score are implemented using the NetworkX library\footnote{https://networkx.org/}, while the others are implemented by raw Python code.
\begin{enumerate}
    \item \textbf{Similarity measures} (number = 4): Jaccard and Cosine distances for followers and followees (all nodes followed by a particular node).  
    \item \textbf{Ranking measure} (number = 2): PageRank \cite{pagerank} for source and destination. This is implemented using the NetworkX library.
    \item \textbf{Graph features} (number = 10): Shortest path between the two nodes (barring edge), belonging to the same community, whether the destination node follows the source node, Adamic-Adar Index \cite{adamicAdar}, Katz Centrality \cite{katz}, HITS Score (Authority and Hubs score). Katz, Authority, and Hubs are calculated for both source and destination nodes.  
    \item \textbf{Follower/followees information} (number = 6): Number of followers and followees of the source and destination each, and the number of common followers and followees.
    \item \textbf{Weight features} from \cite{gbfeatures} (number = 6): The metrics for which these features were generated are - incoming edges to destination (\(w_{in}\)), outgoing edges from source (\(w_{out}\)). In addition to using these two as features, the following 4 features were also derived: 
    \begin{enumerate}
        \item \(w_{in} + w_{out}\)
        \item \(w_{in} * w_{out}\)
        \item \(2*w_{in} + w_{out}\)
        \item \(w_{in} + 2*w_{out}\) 
    \end{enumerate}

    \item \textbf{SVD features} (number = 26): We have performed Singular Value Decomposition on the adjacency matrix of the graph using the number of singular values/vectors \((k)\) = 6. The implementation is using the Scipy
    framework. This yields two unitary matrices - \(U\), representing the left singular vectors and having a shape of \((N, k)\), and \(V\), representing the right singular vectors, having a shape of \((k,N)\). Here, \(N\) represents the number of nodes in the network. The 26 dimensions of features derived from \(U\) and \(V\) are as follows - 
    \begin{enumerate}
    \item 1-6: Row of \(U\) matrix corresponding to source node 
    \item 7-12: Row of \(U\) matrix corresponding to destination node 
    \item 13-18: Column of \(V\) matrix corresponding to source node
    \item 19-24: Column of \(V\) matrix corresponding to destination node
    \item 25: Dot product of row vector corresponding to the source node and row vector corresponding to destination node from \(U\) matrix
    \item 26: Dot product of column vector corresponding to the source node and column vector corresponding to destination node from \(V\) matrix
    \end{enumerate}
    
    \item \textbf{Preferential Attachment features} (number = 2): These are the products of in-degrees and out-degrees of source and destination nodes.

\end{enumerate}

Having obtained \textit{node representations}, using the above methods, the embedding for source and destination nodes of an edge are combined using the Hadamard product to yield each the \textit{edge's representation}. 

\begin{table*}[t!]
    \centering
    \scriptsize
    \begin{tabular}{llrrr}
        \toprule
        \multicolumn{5}{c}{\textbf{Epinions}} \\\cmidrule{1-5}
        \textbf{Category} &\textbf{Algorithm} &{\textbf{XGBoost}} &{\textbf{RandomForest}} &{\textbf{LightGBM}} \\\cmidrule{1-5}
        \textbf{Structural Node Embedding} &\textbf{GraphWave}  &\textbf{0.7922} &0.6927 &0.7905 \\\cmidrule{1-5}
        \multirow{5}{*}{\textbf{Neighborhood Based Embedding}} &\textbf{Node2Vec} &0.6701 &0.6638 &0.6678 \\\cmidrule{2-5}
        &\textbf{DeepWalk} &0.6594 &0.6432 &0.6577 \\\cmidrule{2-5}
        &\textbf{Walklets} &0.7780 &0.7630 &0.7678 \\\cmidrule{2-5}
        &\textbf{NetMF} &0.7654 &0.7602 &0.7608 \\\cmidrule{2-5}
        &\textbf{NodeSketch} &0.8245 &0.8192 &\textbf{0.8223} \\\cmidrule{1-5}
        \multirow{3}{*}{\textbf{Deep Learning Based Methods}} &\textbf{GCN} &0.8877 &0.8884 &0.8883 \\\cmidrule{2-5}
        &\textbf{GraphSAGE} &0.8916 &0.8822 &0.8874 \\\cmidrule{2-5}
        &\textbf{Watch Your Step} &\textbf{0.9106} &0.9103 &0.9090 \\\midrule
        \multicolumn{2}{l}{\textbf{Heuristic Based Features}} &0.9458 &0.9466 & \textbf{0.9470} \\
        \bottomrule
    \end{tabular}
    \captionsetup{justification=centering}
    \caption{Test F1 scores for different categories of algorithms for predicting links in the soc-Epinions network}\label{tab:2}
\end{table*}

\subsubsection{Combinations of different feature paradigms}

Besides using the feature sets generated above, which use either heuristics or representation learning, a new feature set created by combining these two paradigms is introduced. In effect, for each edge, if \(H\) represents the heuristic feature set and \(R\) represents the learned representations (through any of the above-mentioned methods like Structural or GNN-based embeddings), then we have:\\
\begin{equation}\label{eq1}
H = (h_0, h_1, .... , h_{55})
\end{equation}
\begin{equation}\label{eq2}
R = (r_0, r_1, .... , r_{63})
\end{equation} 
And our new combined feature set is a simple concatenation of \ref{eq1} and \ref{eq2}: 
\begin{equation}
f(H,R) = (h_0, h_1, ...., h_{55}, r_0, r_1, ...., r_{63})    
\end{equation}
There are several reasons for choosing concatenation to combine the two feature vectors. Firstly, concatenation keeps information from both constituent feature sets intact. It essentially passes information that is surplus with respect to each individual feature set. This is an advantage when compared with other techniques like addition, averaging, and dot product, all of which condense the information in some form or the other. Secondly, it is possible to leverage the aforementioned advantage because of the relatively smaller dimensionality of the feature sets. The total dimension of the combined feature set is 120, which is not too much higher than either 64 (the dimension of learned representations of graph nodes alone) or 56 (the dimension of heuristics alone). The experiments also confirm the same idea and no substantial increase in training times are observed for this new feature set. Finally, the simplicity of the concatenation operation is advantageous. 

The primary purpose of developing and experimenting with this feature set serves is to explore combining representations from black-box models with completely interpretable hand-engineered features. Additionally, this could also help answer questions like whether using this extra information in the combined feature set can outdo the performance using any of the component paradigms alone. Further, comparing the performance of this combined feature set with the individual feature performances could throw light on how much information overlap is present between the learned representations and the explainable heuristics. Along with that, this could help in studying the extent to which each of the two individual feature sets affects the final prediction performance. 

We carry out this experiment using only a select set of node representation learning algorithms, after excluding those which do not perform significantly well individually. We combine representations learned through GCN, GraphSAGE, and Watch Your Step with the 56 manually engineered heuristic features. 

\subsection{Classification}

The final stage of the methodology is the prediction of links. The features created above for each edge in the network are passed into machine learning and deep learning models to classify whether a link is positive (belongs in the network) or negative (does not belong in the network). As this is modeled as a binary classification task, F1 scores are used as the metric. The machine learning-based models used are LightGBM Classifier \cite{lightgbm}, Random Forest Classifier \cite{randomforest} and XGBoost Classifier \cite{xgboost}. In addition to these, we use custom neural network (NN) architectures for the task of prediction. This includes using towered models where each input tower or leg of the neural network serves to process a single feature set. This single feature set, passing through a single tower of a NN, could either be the isolated source and destination node representations, that get processed by a few NN layers prior to getting combined as a single edge representation, or could be the vector of entire heuristic features or representations learned through some algorithm, that get combined in the deeper layers of the network.

\begin{table*}[h!]
    \centering
    \scriptsize
    \begin{tabular}     {p{0.22\textwidth}p{0.1\textwidth}p{0.04\textwidth}p{0.04\textwidth}p{0.04\textwidth}p{0.04\textwidth}p{0.04\textwidth}p{0.04\textwidth}p{0.04\textwidth}p{0.04\textwidth}p{0.04\textwidth}}\toprule
    \textbf{Algorithm} &\textbf{Model} & \textbf{soc-Epinions} & \textbf{GitHub} & \textbf{Twitch-DE} & \textbf{Twitch-EN} & \textbf{Twitch ES} & \textbf{Twitch-FR} & \textbf{Twitch-PT} & \textbf{Twitch-RU}  & \textbf{wiki-Vote}\\\midrule
    \multirow{3}{*}{\textbf{GCN}} &\textbf{XGBoost} &0.8876 &0.7841 &0.7767 &0.7197 &0.8092 &0.7677 &0.8149 &0.8167 &0.8895 \\ \cmidrule{2-11}
    &\textbf{RandomForest} &0.8883 &0.7788 &0.7768 &0.7188 &0.8004 &0.7664 &0.808 &0.8168 &0.8903 \\ \cmidrule{2-11}
    &\textbf{LightGBM} &0.8882 &0.7801 &0.7727 &0.7229 &0.8061 &0.7639 &0.8105 &0.8179 &0.8900 \\ \midrule
    \multirow{3}{*}{\textbf{GraphSAGE (64)}} &\textbf{XGBoost} &0.8915 &0.8379 &0.8296 &0.7672 &0.8271 &0.8335 &0.8475 &0.8298 &0.8987 \\ \cmidrule{2-11}
    &\textbf{RandomForest} &0.8821 &0.8364 &0.8275 &0.7525 &0.8256 &0.8331 &0.8446 &0.8300 &0.8993 \\ \cmidrule{2-11}
    &\textbf{LightGBM} &0.8873 &0.8378 &0.8306 &0.7652 &0.8275 &0.8349 &0.8470 &0.8295 &0.8991 \\ \midrule
    \multirow{3}{*}{\textbf{WYS}} &\textbf{XGBoost} &0.9105 &0.8195 &0.8145 &0.7667 &0.842 &0.8292 &0.8477 &0.8253 &0.8980 \\ \cmidrule{2-11}
    &\textbf{RandomForest} &0.9102 &0.791 &0.7979 &0.7608 &0.8314 &0.8113 &0.8404 &0.8185 &0.8914 \\ \cmidrule{2-11}
    &\textbf{LightGBM} &0.9090 &0.8184 &0.8113 &0.7626 &0.8405 &0.8272 &0.8502 &0.8233 &0.8953 \\ \midrule
    \multirow{3}{*}{\textbf{Heuristic}} &\textbf{XGBoost} &0.9458 &0.8361 &0.8818 &0.7485 &0.8779 &0.8871 &0.9054 &0.8414 &0.9665 \\ \cmidrule{2-11}
    &\textbf{RandomForest} &0.9465 &0.8543 &0.8858 &0.7794 &0.8802 &0.8886 &0.9089 &0.8615 &0.9604 \\ \cmidrule{2-11}
    &\textbf{LightGBM} &0.9469 &0.8412 &0.8831 &0.7443 &0.8791 &0.8897 &0.9055 &0.8443 &0.9676 \\ \midrule
    \multirow{3}{*}{\textbf{GCN + Heuristic}} &\textbf{XGBoost} &0.9458 &0.8356 &0.8815 &0.7479 &0.8775 &0.8888 &0.9067 &0.8418 &0.9667 \\ \cmidrule{2-11}
    &\textbf{RandomForest} &0.9410 &\textbf{0.8629} &\textbf{0.8879} &0.7860 &0.8842 &0.8914 &\textbf{0.9099} &0.8660 &0.9575 \\ \cmidrule{2-11}
    &\textbf{LightGBM} &0.9473 &0.8416 &0.8835 &0.7448 &0.8783 &0.8898 &0.9055 &0.8426 &\textbf{0.9679} \\ \midrule
    \multirow{3}{*}{\textbf{GraphSAGE (64) + Heuristic}} &\textbf{XGBoost} &0.9459 &0.8353 &0.8822 &0.7491 &0.8787 &0.8888 &0.9045 &0.8426 &0.9661 \\ \cmidrule{2-11}
    &\textbf{RandomForest} &0.9429 &0.8623 &0.8868 &0.7882 &0.8817 &0.8916 &0.9075 &0.8627 &0.9560 \\ \cmidrule{2-11}
    &\textbf{LightGBM} &0.9473 &0.8415 &0.8831 &0.7482 &0.8800 &0.8909 &0.9066 &0.8454 &0.9671 \\ \midrule
    \multirow{3}{*}{\textbf{WYS + Heuristic}} &\textbf{XGBoost} &0.9462 &0.8416 &0.8842 &0.7526 &0.8795 &0.8901 &0.9037 &0.8439 &0.9661 \\ \cmidrule{2-11}
    &\textbf{RandomForest} &0.9463 &0.8625 &0.8868 &\textbf{0.7924} &\textbf{0.8846} &0.8913 &\textbf{0.9099} &\textbf{0.8644} &0.9599 \\ \cmidrule{2-11}
    &\textbf{LightGBM} &\textbf{\textbf{\textbf{\textbf{0.9479}}}} &0.8486 &0.8834 &0.7477 &0.8808 &\textbf{0.8917} &0.9046 &0.8467 &0.9667 \\ 
    \bottomrule
    \end{tabular}
    \captionsetup{justification=centering}
    \caption{Test F1 scores for link prediction using all categories of algorithms and features by combining Heuristics and Learned Representations. }\label{tab:3}
\end{table*}

\begin{table*}[!h]
    \small
	\centering
	\begin{threeparttable}
	\begin{tabular}{p{0.08\textwidth}p{0.07\textwidth}p{0.3\textwidth}p{0.06\textwidth}p{0.06\textwidth}p{0.06\textwidth}p{0.06\textwidth}p{0.06\textwidth}}
	    \toprule
	   \textbf{Input Dims} &\textbf{Input} &\textbf{Combination method} &\textbf{Params} &\textbf{Activ.} &\textbf{Epinions } &\textbf{Github} &\textbf{wiki-vote} \\\midrule
        64 & \(S, D\) & \(e(S,D)\) & 84K & ReLU & 0.9097 & 0.8189 & 0.8891 \\
        
        56 & \(H\) & NA & 123K & ReLU & 0.9530 & 0.8906 & 0.9434 \\
        
        64+56 & \(S, D, H\) & \(H  \|  e(S,D)\) & 236K & ReLU & \textbf{0.9532} & 0.8706 & 0.9471 \\
        
        64+64 & \(S, D\) & \(e(f_4(S),f_4(D))\) & 167K & ReLU & 0.9118 & 0.8792 & 0.9526 \\
        
        64+64+56 & \(S, D, H\) & \(e(f_4(H),f_4(S),f_4(D))\) & 249K & ReLU & 0.9411 & \textbf{0.8923} & 0.9595 \\
        
        64+64+56 & \(S, D, H\) & \(e(f_4(H), f_2(e(f_4(S), f_4(D))))\)  & 197K & ReLU & 0.9432 & 0.8599 & 0.9589 \\
        
        64+64 & \(S, D\) & \(f_3(S) \| f_3(D)\) & 193K & ELU & 0.9107 & 0.8772 & 0.9511 \\
        
        56 & \(H\) & NA & 18K & ELU & 0.9451 & 0.8881 & 0.9469 \\
        
        64+64+56 & \(S, D, H\) & \(H \| f_3(S) \| f_3(D)\) & 200K & ELU & 0.9526 & \text{0.8903} & 0.9572 \\
        
        64+64 & \(S, D\) & \(e(f_3(S), f_3(D))\) & 160K & ELU & 0.9135 & 0.8783 & 0.9538 \\
        
        64+64+56 & \(S, D, H\) & \(e(f_1(H), f_2(e(f_4(S), f_4(D))))\) & 168K & ELU & 0.9500 & 0.8852 & \textbf{0.9607} \\
	   \bottomrule
	\end{tabular}
	\begin{tablenotes}
        \item[1] \(e:\) Hadamard/element-wise product of vectors. 
        \item[2] \(f_n:\) Passing the input through \(n\) fully connected layers.
        \item[3] \(\| : \) Horizontal concatenation of vectors.
        \item[4] \(S =\) WYS embedding of source node. 
        \item[5] \(D =\) WYS embedding of destination node. 
        \item[6] \(H =\) Heuristic-based features.
    \end{tablenotes}
    \end{threeparttable}
    \caption{Test F1 scores of performing link prediction using custom Neural Network architectures. A description of the architecture is also presented.}
    \label{tab:4}
\end{table*}

\section{Experimental results}
\label{ER}
These experiments focus on comparing the performances of the aforementioned feature engineering techniques, in the link prediction task. As the primary focus for this task is to identify the best technique for generating edge representations, and not the models for classification, extensive hyper-parameter tuning was not carried out for each triplet of a network, a feature generation algorithm, and a model. The set of hyperparameters values for each model is chosen after observing the performance with a variety of model specifications on the Epinions data set. The selected specifications can be found in our code.

\subsection{Initial Comparison of Algorithms}
The first experiment is to test the performance of all algorithms individually on the Epinions data set. The first runs are performed on this data set due to its popularity in studies pertaining to social networks. Table \ref{tab:2} contains the testing F1 scores given by each algorithm-model combination. The highlighted values are the best scores in each category of algorithms.

These findings suggest that the heuristic-based features are the most effective. The neighborhood-based embedding methods perform considerably worse than all other methods, particularly the GNN-based and Heuristic features. 

\subsection{Multiple Data Sets}
Keeping with the observations seen in the previous experiment, the better-performing algorithms are tried on different data sets with different properties. For all of these, experiments are also conducted with combinations of the heuristic-based features and each of the embeddings generated using the deep learning techniques. The results are shown in Table \ref{tab:3}.

It is evident that the best performance is observed for a combination of Heuristic-based features with a learned representation of nodes. The combination with Watch Your Step gives the best results for 6 out of 9 data sets and the combination with GCN gives the best performance with 4 out of 9 data sets (the F1 score is the same for Twitch-PT). As far as the models are concerned, Random Forest is the best model for 6 out of 9 data sets and LightGBM is the best model for the rest. This certainly points in the direction of black-box representations and heuristics supplementing each other and enhancing the overall performance. However, as the best test F1 scores, obtained through a combination of heuristic features and node representations, improve only slightly upon the performance with heuristics alone, it may indicate that overlapping information is present in the two different feature sets.

\begin{figure*}[h]
    \centering
    \includegraphics[width=0.48\textwidth]{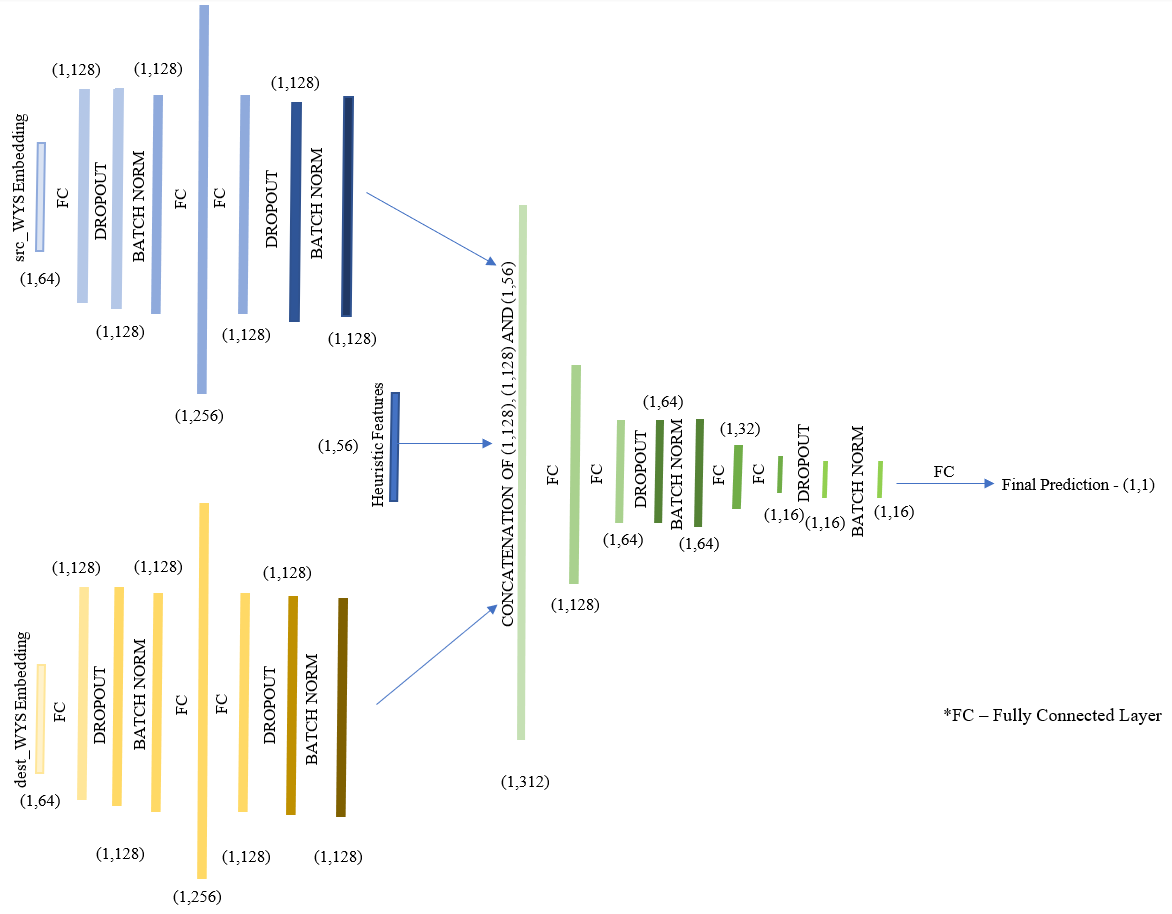}
    \caption{A three-tower Neural Network Model as used in the 9th experiment. The layers depicted in blue process the WYS embedding for the source node, and the layers depicted in yellow process the WYS embedding for the destination node. These get combined with Heuristic features in the deeper layers of the network (depicted in green). Although this is not the best model for either data set, it performs well for all.}
    \label{fig:2}
\end{figure*}

\subsection{Deep Learning Architectures}
It is apparent that machine learning models give the best results with a deterministic set of interpretable features. Considering the possibility that the information learned by neural-network-based representation learning algorithms might not be fully captured by models using bagging and boosting of Decision Trees, we try various custom deep-learning architectures for the task of binary classification. A description of the models used along with the results on them is shown in Table \ref{tab:4}. The different structures and hyper-parameters were arrived at empirically. The reason why only Epinions, GitHub, and the wiki-Vote datasets were used for this experiment, is the substantially smaller size of the remaining Twitch networks. As they have a much lower number of both nodes and edges, in keeping with the general expectation that neural network models would not perform better on smaller datasets, they are excluded. Figure \ref{fig:2} portrays one of the network architectures used. This is a model that performs consistently well on all three datasets. 

As can be observed, some of these architectures are capable of outperforming tree-based ensemble models. For all three data sets, multi-input neural-network architectures that take in Watch Your Step representations for the source and destination nodes along with Heuristic-based features work best. For Epinions and GitHub, the performance significantly exceeds that given by any machine learning model. 

\section{Conclusion}
\label{Conc}
It is clear that the most effective singular method to generate edge features for the task of link prediction is by using deterministic metrics describing the characteristics of nodes, edges, and graphs. However, as has been observed across all data sets, the performance of a combination of Heuristic-based features and node representations learned using a deep learning algorithm enhances performance, albeit slightly. This goes to prove that a combination of features from different paradigms can augment each other. It also inspires further study into how important each feature set might be, and whether their effectiveness is correlated with interpretability. The experiments also show that using custom feed-forward deep neural networks for the task of classification of the edges proves to be better than traditional machine learning models. 

\section{Future Scope}
\label{FS}
The experimental results encourage a deeper exploration of ways to combine different types of information to predict human behavior on social media platforms. The first step in that direction could be to identify the intrinsic aspects captured by different types of features. This could involve using features based on more rounded information about the users, or understanding what exactly representation learning models capture and how in each feature dimension. An exhaustive experimental setup that explores more combination methods, possibly even combining learned representations from dissimilar pairs of algorithms, could lead to interesting insights. Finally, the findings from the experiments in this work could be leveraged to create a holistic framework that recommends social media connections using the predictions, followed by ranking. 

Another, parallel line of analysis could be the effect of the combination-based strategy of different types and sizes of networks. As observed with Twitch, although the networks corresponding to different countries represent the same type of relationship, the F1 scores are very different. This could be due to a relatively small amount of data present for each graph or due to real differences in the characteristics of the graphs stemming from cultural diversity. In either case, analysis of the given technique on a wider selection of data sets would yield a better understanding of what would constitute an effective technique for the task of link prediction.

\bibliographystyle{named}
\bibliography{ijcai23}

\end{document}